\documentclass{sig-alternate}

\usepackage{endnotes}
\usepackage{times}
\usepackage{balance}
\usepackage{flushend}
\usepackage{amsfonts}
\usepackage{amssymb}
\usepackage{amscd}
\usepackage{amsmath}
\usepackage{array}
\usepackage{epstopdf}
\usepackage{multirow}
\usepackage{graphicx}
\usepackage{subfigure}
\usepackage{url}
\usepackage[
  breaklinks=true,
  unicode=true,
  urlcolor = blue,
  colorlinks = true,
  citecolor = blue,
  linkcolor = blue]{hyperref}

\makeatletter
\g@addto@macro{\UrlBreaks}{\UrlOrds}
\makeatother
\usepackage[hyphenbreaks]{breakurl}
\usepackage{tabularx}
\usepackage{algorithm,algpseudocode}
\usepackage{algorithmicx}
\usepackage{hhline}

\usepackage{slashbox}
\usepackage{bm}

\newcommand{\ignore}[1]{}
\newcommand{\revised}[1]{}
\newcommand\comment[1]{}

\newcommand{\tabincell}[2]{\begin{tabular}{@{}#1@{}}#2\end{tabular}}

\usepackage{color}
\usepackage{listings}
\usepackage{soul}
\usepackage{graphicx}
\begin{document}

\title{An Empirical Study on Android for Saving Non-shared Data on Public Storage
\titlenote{All vulnerabilities described in this paper have been reported to corresponding companies. Alipay has fixed this vulnerability in the latest version. We have got the IRB approval before all experiments related to human subjects.}}

\numberofauthors{3}
\author{
\alignauthor Xiangyu Liu, Zhe Zhou, Wenrui Diao\\
       \affaddr{Department of Information Engineering} \\
       \affaddr{The Chinese University of Hong Kong}
\alignauthor Zhou Li\\
       \affaddr{RSA Laboratories}\\
       \affaddr{11 Cambridge Center \\Cambridge, MA 02142 USA}
\alignauthor Kehuan Zhang\\
       \affaddr{Department of Information Engineering} \\
       \affaddr{The Chinese University of Hong Kong}
}

\maketitle

\begin{abstract}

With millions of apps that can be downloaded from official or third-party market, Android has become one of the most popular mobile platforms today. These apps help people in all kinds of ways and thus have access to lots of user's data that in general fall into three categories: sensitive data, data to be shared with other apps, and non-sensitive data not to be shared with others. For the first and second type of data, Android has provided very good storage models: an app's private sensitive data are saved to its private folder that can only be access by the app itself, and the data to be shared are saved to public storage (either the external SD card or the emulated SD card area on internal FLASH memory). But for the last type, i.e., an app's non-sensitive and non-shared data, we found that there is a big problem in Android's current storage model which essentially encourages an app to save its non-sensitive data to shared public storage that can be access freely by all other apps.

At first glance, it seems no problem to do so, as those data are non-sensitive after all, but it implicitly assumes that the app developers could correctly identify all sensitive data and prevent all possible information leakage from private-but-non-sensitive data. In this paper, we will demonstrate that this is an invalid assumption with a thorough survey on information leaks of those apps that had followed Android's recommended storage model for non-sensitive data.
Our studies showed that highly sensitive information from billions of users can be easily hacked by exploiting the mentioned problematic storage model. Although our empirical studies are based on a limited set of apps, the identified problems are never isolated or accidental bugs of those apps being investigated. On the contrary, the problem is rooted from the vulnerable storage model recommended by Android (even including the latest Android 4.4 at the time of writing). To mitigate the threat, we also propose a defense framework.
\end{abstract}

\section{introduction}
\label{attest:intro}

The last decade has seen the immerse evolvement of smartphone technologies. Today's smartphones carry much more functionalities than plain phones, including email checking, social networking, gaming and entertainment. These emerging functionalities are largely supported by the vast amount of mobile applications (\textit{apps}). As reported by~\cite{appbrain}, the number of Android apps on Google Play is hitting 137 million.

When apps are performing their tasks, it is inevitable for them to touch our private information, like emails, contacts, various accounts, etc. Generally, when people download and install an app, they trust it and hope it could protect their private information carefully. To support the protection of private data, Android system allocated for each app a private folder that can only be accessed by the owner app.

Since each app now has it own private folder, it seems natural and absolutely right to save all private data that should not be shared with others to this secure folder, but unfortunately, according to our survey to be presented in section~\ref{sec:survey}, this is not the case. We found that, for private data, many apps will first try to decide what are sensitive and what are not, then only store the identified sensitive private data to the secure private folder, and save the other non-sensitive private data to a ``private'' folder on shared public storage, even though the data should never be shared with other apps.

Although there are some reasons and benefits by differentiating sensitive private data from non-sensitive private data and save the latter to shared public storage, which will be discussed in section~\ref{sec:discussion}, we believe that such practice is wrong and could lead to serious attacks. \textit{App's private data, which are not supposed to be shared with other apps, should never be saved to shared public storage, no matter the private data is identified as sensitive or not}. Since most app developers are not experts in security and privacy, thus could make wrong decisions on what are sensitive and what are not. According to our survey in Section~\ref{sec:survey}, we indeed have found some very popular apps leak information that are obvious sensitive (like full name, phone numbers, and etc), which shows even the companies with highly reputation are not so credible under such a wrong storage model. What's worse, even all developers have made the right decisions on each piece of information generated by their own apps, it would still be problematic, because non-sensitive information from multiple apps/sources could become sensitive and lead to serious privacy attacks. In section~\ref{sec:survey}, we will give examples of information that is not sensitive at first glance, but has huge privacy impact when combined with other public information. Some examples described in Section~\ref{sec:attack} demonstrate how such non-obvious sensitive information could be used in serious privacy attacks.

Surprisingly, the storage model is still recommended in official Android documents, and the latest Android 4.4 emphasized a misleading concept called ``app-private directory''. Following statement is quoted from official Android developer documents~\footnote{\url{http://developer.android.com/guide/topics/data/data-storage.html}}: \textit{
  ``If you are handling files that are not intended for other apps to use, you should use a {\bf private storage directory} on the {\bf external storage} by calling getExternalFilesDir()''}. According to the same page, the {\bf external storage} is defined as a place to \textit{``store public data on the shared external storage''}. So, this guideline is actually asking apps to save app-private data to shared public storage. More relevant details on storage security in latest Android 4.4 will be given in section~\ref{androidStorageModel}.

Although it has been a hot topic to sniff and infer sensitive information from Android devices and there are a number of researches demonstrating how to launch attacks successfully, the work presented in this paper is different. Here, we want to report a serious problem of wrong private data storage model that has been followed by many apps for years, while the previous works either trick the user to grant high-profile permissions~\cite{bugiel2012towards} or exploit the system flaws in side-channel attacks~\cite{jana2012memento}. It is also worth noting that the attacks proposed in this paper are new and different from attacks have been reported, even if the target is the same app. For example, our second attack against \texttt{WhatsApp} is based on non-obvious sensitive data (photo names), while the attack reported \cite{whatsappuserchat} took \texttt{WhatsApp} database as the target, and our attack succeeds in its latest version, even after the reported vulnerability was fixed.

Since the reported problem goes beyond the vulnerability of any specific app, current data protection mechanisms, like data encryption, do not work well. To address this problem, developers may be required to update their codes by saving all private data to the securely protected private folder instead of the one located on shared public storage. However, expecting all the developers to obey the security rules and fix this security issue is impractical. Another more effective solution requires some updates to Android system by enforcing fine-grained access control on ``app-private'' data. In order to mitigate this problem as much as possible, while also avoid bringing unnecessary trouble to users, we propose to argument the existing security framework on public storage by instrumenting the original APIs and automatically checking the ownership of files and directories.

\textbf{Our Contributions}. We summarize our contributions as follows:
\begin{itemize}

\item We revisit the Android data storage model and first study 17 popular Android apps on whether they store the data correctly. Our study reveals that 13 apps (billions installations on smartphones in total\footnote{The detailed statistical data are shown in Table \ref{privacyevaluation} and Figure \ref{fig:17apps}, which actually contain installations on other platforms except Android, but Android have a big advantage in the size of these users.}) leave user's private information on shared storage, which lead to highly privacy risks. Then, a large scale analysis conducted shows such a roblem is widely exist among apps, which means it is not a bug happens on several apps but a wrong storage model provided for developers.

\item We design three concrete attacks based on the sensitive information collected from the popular apps, including inferring user's location (\texttt{WeChat}), identifying the owners of phone numbers acquired from \texttt{WhatsApp}, and inferring the chatting buddy (\texttt{KakaoTalk}). These attacks are new and achieve high accuracy while also remain low-profile.

\item We discuss the feasibility of the existing solutions on protecting data in shared public storage and propose our mitigation approach to protect app-private data. What's more, we also discuss the fundamental reasons lies behind the coarse-grained access control mechanism on shared public storage in Section \ref{sec:discussion}.
\end{itemize}

\textbf{Roadmap}. The rest of the paper is organized as follows: We begin with the adversary model taken by this paper -- Section \ref{adversarymodel}, as well as an overview of Android storage model in Section \ref{androidStorageModel}. Section \ref{sec:survey} shows our surveys on private information leakage of Android apps. Three concrete attacks will be introduced in Section \ref{sec:attack}. Section \ref{sec:defense} is devoted to corresponding defense technologies, and in Section \ref{sec:discussion}, we will discuss the fundamental reasons lead to this problem and also some limitations of this paper. We compare our work with prior related research in Section \ref{sec:relatedWork}. Finally, Section \ref{sec:conclusion} concludes this paper.

\begin{figure}[htb]
    \centering{
		\includegraphics[scale=0.34]{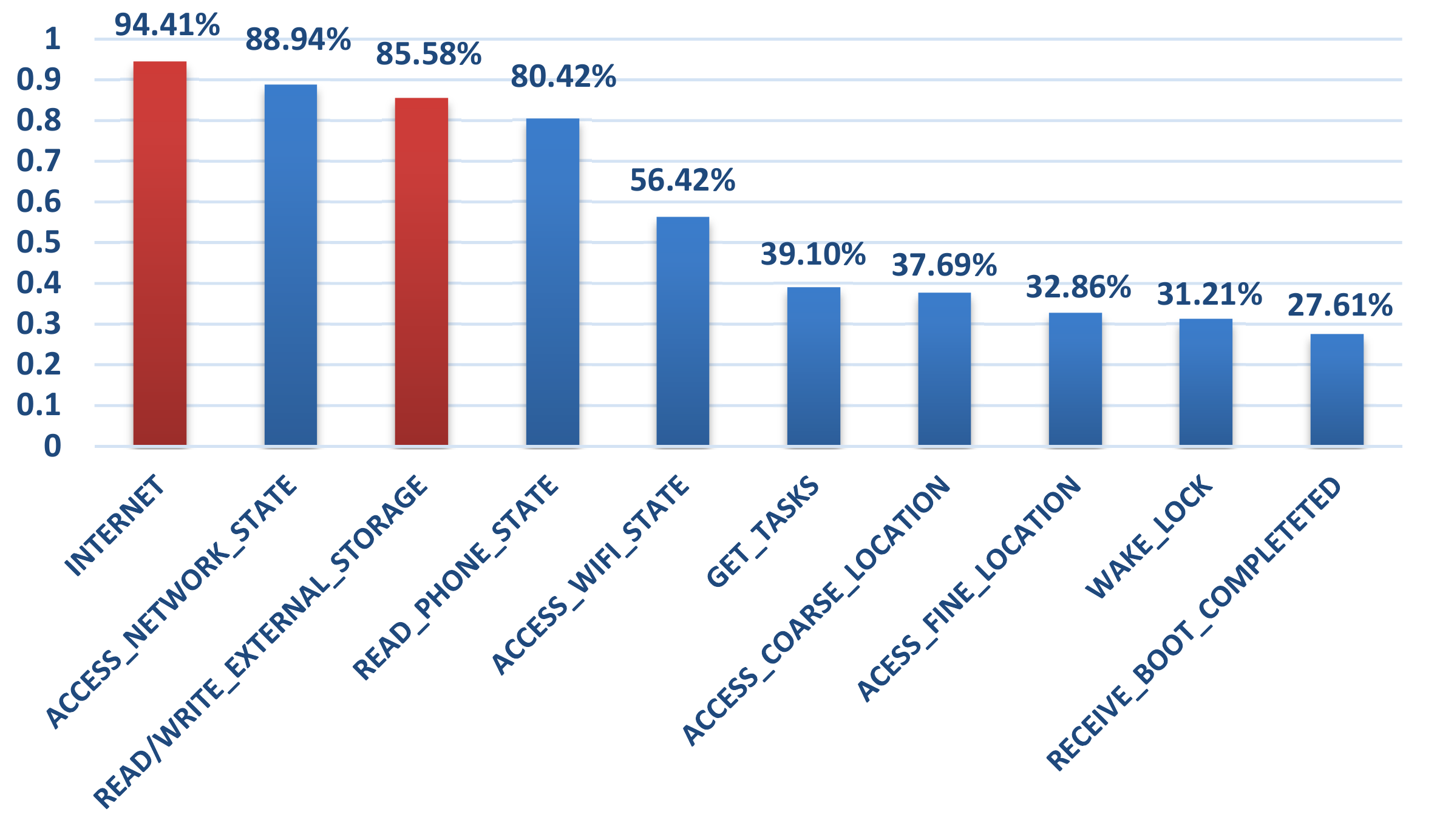}\\
		\caption{Top 10 Permissions Requested by 34369 Apps}
		\label{fig:permission1}}
\end{figure}

\section{Adversary Model}
\label{adversarymodel}

The adversary studied in our paper is interested in privacy attacks by only exploiting the app-private data located on shared public storage due to the wrong model introduced in this paper. In order to get app-private data and perform our attacks, it is assumed that a malicious app with access to external storage (i.e., the \texttt{READ\_EXTERNAL\_STORAGE} or \texttt{WRITE\_EXTERNAL\_STORAGE} permission, \texttt{READ} and \texttt{WRITE} for short) and the Internet (i.e., the \texttt{INTERNET} permission) has been installed successfully on an Android phone updated to latest version (Android 4.4). The malicious app will read certain app-private folders selectively on shared public storage, searching for information that can be used for privacy attack, and then upload such information to a malicious server where the intensive data analysis and concrete attacks will be done. Note that there is no need to scan the whole shared storage or upload all app-private data, because it is reasonable to assume that the adversary has studied many apps in the market beforehand and already know where to read and how to extract useful information from app-private data.

Based on above assumptions, it is hard to detect the attack. First, it requires only two very common permissions. We did a statistical analysis on the permissions requested by 34369 apps, the top 10 permissions are shown in the Figure \ref{fig:permission1}. About 94\% of apps request \texttt{INTERNET} permission, and about 85\% of apps request the permission to access external storage, ranking No.1 and No.3 respectively. Secondly, it only uploads data when Wi-Fi is available and filter out non-useful app-private data to minimize data size. Third, all network- and CPU-intensive operations are done on a remote malicious server instead of the mobile phone, which can eliminate the alarm in CPU and power usage. Finally, the malicious app will not take any other malicious behavior to help the attack, like exploiting other system vulnerabilities or bypassing existing security mechanisms, so it is hard to be tagged as malicious by current anti-virus software. The victim of such attack could be the owner of the compromised Android phone. However, depends on on the apps installed and the information leaked through app-private data, attacks can be extended to user's family members, friends and colleagues.

\section{Overview of Android Storage \\Model}
\label{androidStorageModel}

\begin{figure}[t]
    \centering{
		\includegraphics[width=0.47\textwidth]{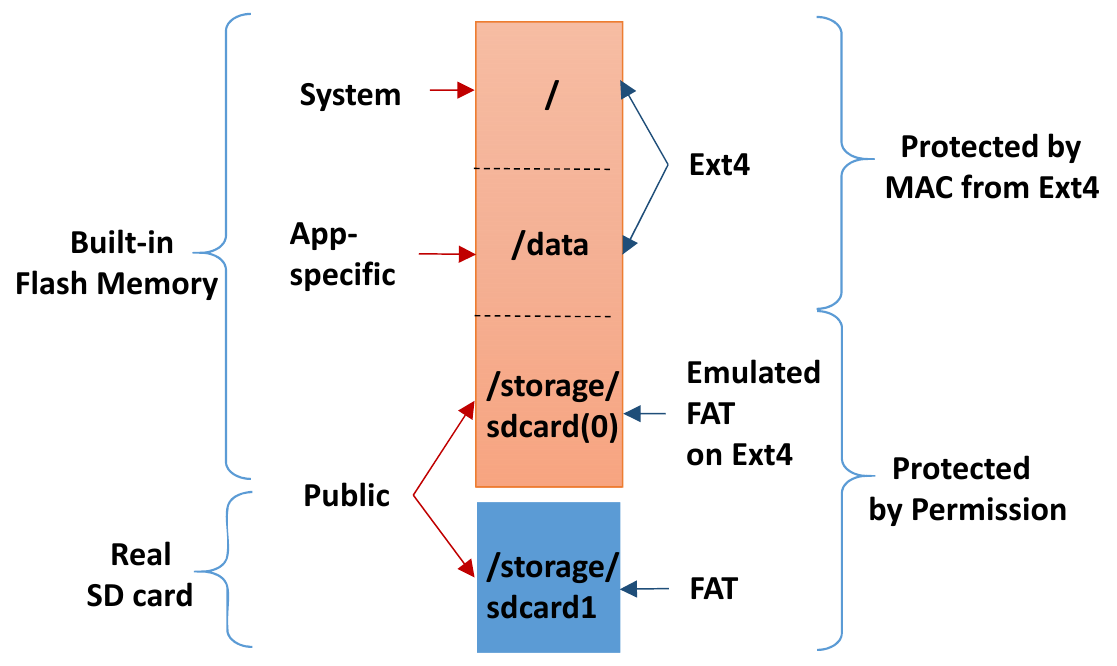}
    \caption{Overview of Android storage options and security}
		\label{fig:storagemodel}}
\end{figure}

In Android system, there are several different storage options which, according to their access control mechanisms, can be divided into three categories: \textit{system}, \textit{app-specific}, and \textit{public}, as shown in Figure~\ref{fig:storagemodel}. The system storage is the directory where the whole Android OS locates, and protected by Linux access control mechanism. The app-specific storages are places that are under control of a specific app (i.e., the owner of these folders), and can only be read and written by that app. Android system provides three types of app-specific storage: \textit{internal storage}~\footnote{The name ``internal storage'' is actually very confusing. It should be understood as ``being accessed only by app internal code'' instead of the physical location of the storage (i.e., build-in Flash memory), because Android also uses a term ``external storage'' that is shared public storage and could be either on build-in Flash-memory or SD card.},  \textit{shared preferences}, and \textit{SQLite databases}. The last storage option is shared public storage which is used to share data among apps, like downloaded documents, video, photos, etc.

\ignore{can only be read and/or write by a system process.  /proc files belong to system? but can be read A typical example}

As shown in Figure~\ref{fig:storagemodel}, for system and app-specific storage, Android relies on the Discretionary Access Control mechanism provided by underlying Linux Ext2/4 file system to enforce access control. Android create an account for each app (UID) and system process which would be the owner of a specific folder and controls who can access that folder. However, there is no fine grained access control on shared public storage which is only protected with \texttt{READ} and \texttt{WRITE} permissions. As a result, an app can read or write any folder at any time once it has acquired corresponding permissions. The Android system even sets the permissions to \texttt{rwxrwx----} for the shared public folders on built-in Flash memory that were originally protected by Ext2/4 with fine grained access control mechanism, which we believe is to emulate the behavior of FAT file system and be consistent with the one used on SD card.

\textbf{Misleading of new features on Android 4.4}. There are several security enhancements on the shared public storage in the latest Android 4.4, which could give people a false sense that Android 4.4 is immune from the attacks introduced here. First, it emphasized the concept ``app-private directory'', that are not intended for other apps to use, and app reads or writes files in its own private directory does not require \texttt{READ} or \texttt{WRITE} permissions~\cite{androiddocumentexternal}. At first glance, many people, even including some security experts, may think that the Android system will save the app-private data in secure places, and the problem discussed in this paper is gone. However, that is totally wrong! The Android system actually creates a folder (named ``./android/data/'') on shared public storage to save all app-private data. As mentioned above, there is no fine grained access control on shared public storage, any app with appropriate permission thus can access those app-private data and apply privacy attacks.

Another misleading feature introduced in latest Android 4.4 is the permission \texttt{WRITE\_MEDIA\_STORAGE}. This permission controls the write operation on \textbf{real} SD card and will only be granted to system process. It can be taken as a non-public permission, which means normal apps could not write to \textbf{real} SD card.  However, there is no dedicated permission to control the read operation on SD card, which makes the latest Android version still vulnerable to the attack on the wrong app-private storage model. This has been confirmed by our home-brewed apps and third-party apps downloaded from Google Play market.

\section{Survey on information leaks \\from shared storage}
\label{sec:survey}

It is true that some app-private data are not sensitive. However, the model of letting app write their private data to shared public storage relies on a strong assumption that app developers can made right decision to tell sensitive data from non-sensitive ones. In this Section, we will present a survey of the information leaks through the app-private data saved on shared public storage, which will show that such assumption is problematic. The survey includes two parts: the first is a detailed examination of information in ``shared'' app-private data for 17 popular apps (in Section~\ref{findingpopular}), and the other is a more general and large scale study on the ``shared'' app-private data on public storage (in Section~\ref{investigationlargescale}).

%\label{sec:attack1}
\subsection{Investigation on popular apps}
\label{findingpopular}

\noindent\textbf{What apps have been surveyed}. According to our adversary model, the attackers are interested in privacy attacks over ``shared'' app-private data, so we have selected 17 most popular apps on Google Play from three categories: ``social networking'', ``instant messaging'' and ``online payment'', which are believed to be more likely to touch users' sensitive information. The apps, including their user numbers, are shown in Figure~\ref{fig:17apps}, while the versions of these apps installed on our mobile phones are list in the Appendix~\ref{appendix1}.

\noindent\textbf{How to check the app-private data}. These 17 apps are downloaded and installed on all three Samsung Galaxy S3 mobile phones. Then we manually simulate three different users on three phones respectively, including account registration, adding good friends, sending message, etc. Finally, we check the shared public storage, search sensitive data for each app, and classify the collected data.

\begin{figure*}[ht]
    \centering{
		\includegraphics[scale=0.85]{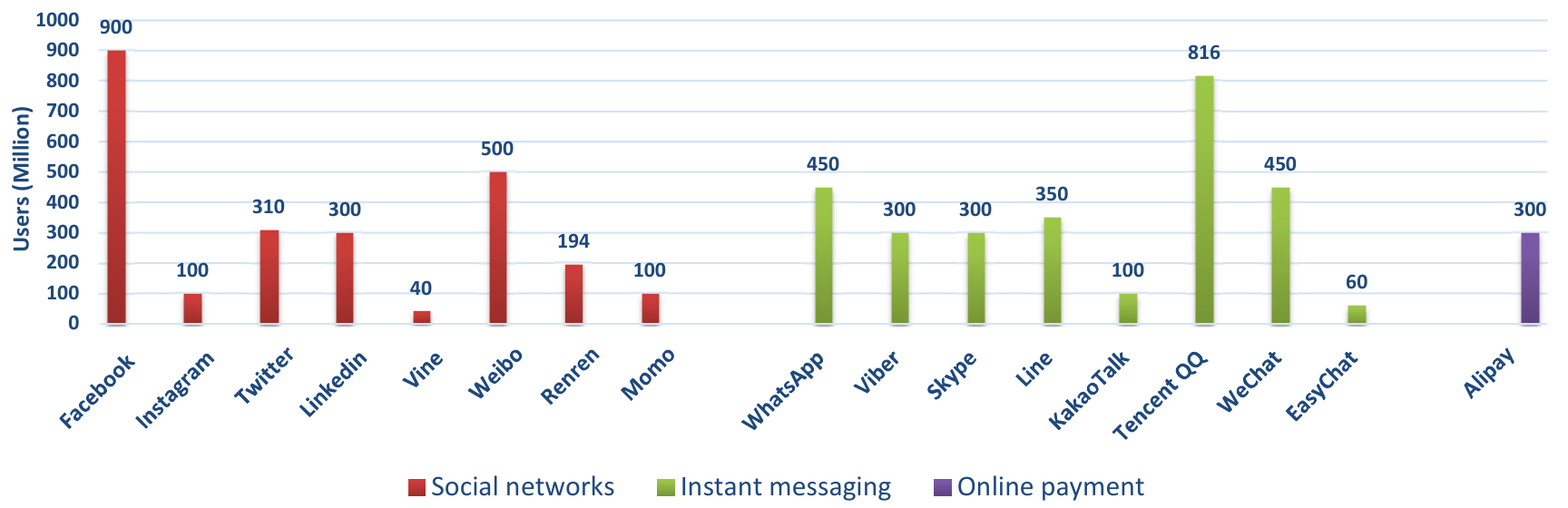}
	\caption{The number of users installed the 17 popular apps (Details are shown in the Appendix with citations)}
		\label{fig:17apps}}

\end{figure*}

\begin{table}[htb]
\centering
\caption{Sensitive data acquired from Smartphone A}
\newsavebox{\tableboxy}
\label{privacycontent}
\begin{lrbox}{\tableboxy}
\begin{tabular}{l|l|l} \hline

\textbf{\tabincell{l}{Sensitive \\Information}}& \textbf{App Name}& \textbf{Content/Remarks}\\ \hline

\multirow{5}{*}{\textbf{User Identity}} & Viber & \tabincell{l}{Name} \\ \hhline{~--}
     & Renren & UID\\ \hhline{~--}
     & Weibo &  UID\\ \hhline{~--}
     & Linkedin & Profile photo \\ \hline

\multirow{3}{*}{\textbf{Phone Number}} & Viber & 8525984xxxx \\ \hhline{~--}

    & Alipay & 1521944xxxx  \\ \hhline{~--}

    & EasyChat & 8525984xxxx  \\\hline

\multirow{2}{*}{\textbf{Email}} & Weibo & lixxxxxxx@163.com \\ \hhline{~--}

    & Renren$^{\star}$ & \tabincell{l}{lixxxxxxx@gmail.com} \\ \hline

\multirow{5}{*}{\textbf{Account}} & Tencent QQ & 8387xxxxx \\ \hhline{~--}
%    & Facebook & Email/username$^{\star}$ \\ \hhline{~--}
    & Renren & \tabincell{l}{2388xxxxx}  \\ \hhline{~--}
    & Momo & 3120xxxx  \\ \hhline{~--}
    & Weibo & 2648xxxxxx \\ \hhline{~--}
    & WeChat & \tabincell{l}{QQ account / Phone number} \\ \hline

\multirow{5}{*}{\textbf{Connection}} & EasyChat & \tabincell{l}{Call Records} \\ \hhline{~--}
    & Linkedin & \tabincell{l}{Profile photos} \\ \hhline{~--}

    & KakaoTalk &\tabincell{l}{Names of content directories}\\ \hhline{~--}

& WhatsApp & \tabincell{l}{Phone Numbers of Friends:\\1213572xxxx ...}\\ \hhline{~--}

& Renren & \tabincell{l}{Accounts of Friends:\\8295xxxxx ...}\\ \hline

\end{tabular}
\end{lrbox}
\scalebox{0.8}{\usebox{\tableboxy}}
\\ $^{\star}$ The Email found by grep command in Table~\ref{privacycontent} is from a ``log.txt'' file left by Renren old version (\texttt{5.9.4}).
\end{table}

\noindent\textbf{How the information is leaked}. By studying the 17 sample apps, we found 10 of them leak various user-related information through app-private data on shared public storage, including full name, account name and emails, etc., as shown in Table~\ref{privacycontent}. Such information is leaked in different forms which will discussed below, and the details of extraction is elaborated in Appendix~\ref{appendix2}.

\vspace{2pt}\noindent \textit{Leak through text file:} Some apps store user's profile into a text file. For example, \texttt{Viber} directly saves user's real name, phone number and path of profile photo into a plain-text file \texttt{$\sim$/viber/\\.viber/.userdata} without any encryption ($\sim$ represents the root directory of public storage). User's username, email address\footnote{If the user uses email to register Weibo account. The user also can use phone number to register an account.} are stored by \texttt{Weibo} in a file named by the user's UID. Some apps keep text logs which also reveal quite rich information. \texttt{EasyChat} keeps call records in a file \texttt{$\sim$/Yixin/log/pjsip\_log.txt}, so caller's number, callee's number and call duration can be easily recovered by simply parsing each record.

\begin{figure}[tb]
    \centering{
		\includegraphics[scale=0.17]{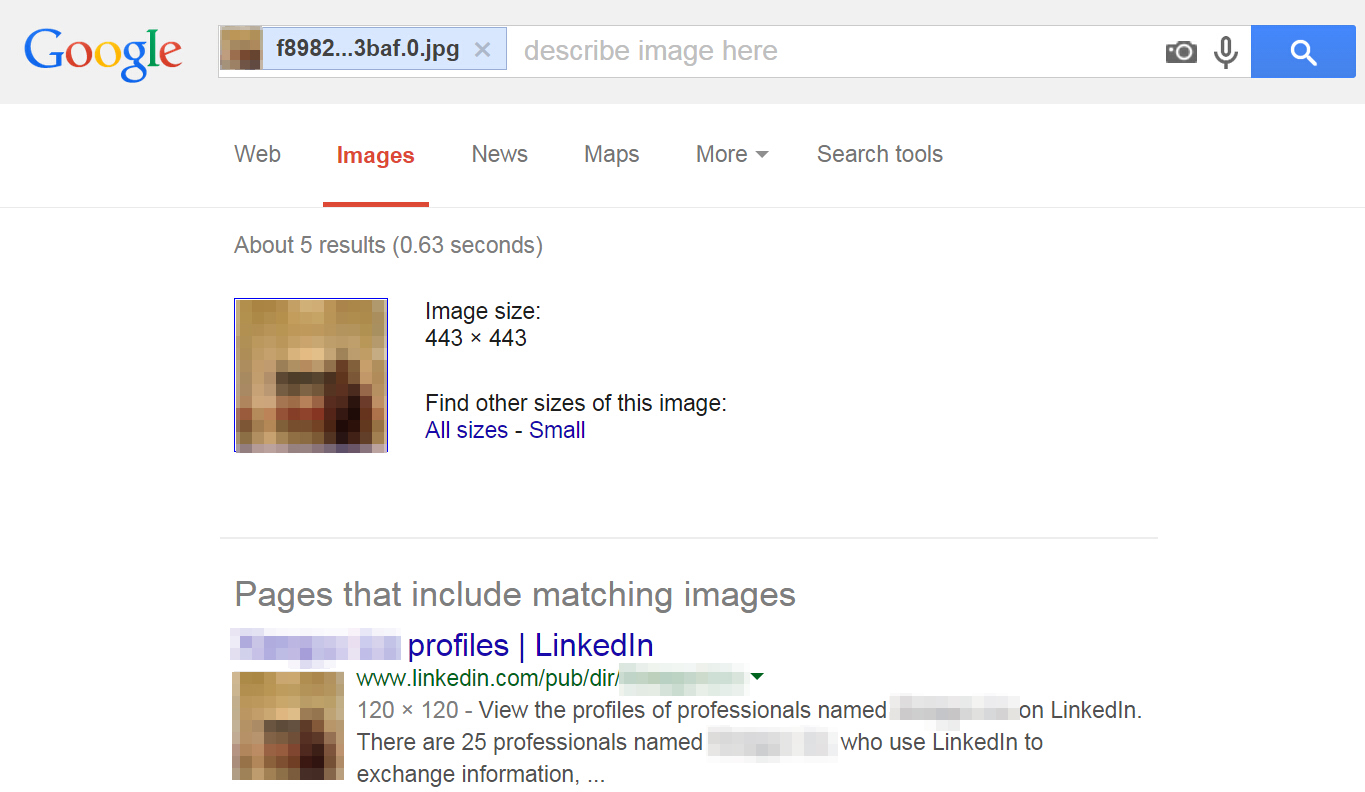}
	\caption{The result of searching a user's Linkedin profile photo}
		\label{fig:linkedin}}
\end{figure}

\vspace{2pt}\noindent \textit{Leak through photo:} The social network apps usually cache user profile photos in shared public storage, like \texttt{LinkedIn} in our studies. The photo itself is non-obvious sensitive information, since we can easily get millions of such photos with Google. However, our study shows user's \texttt{LinkedIn} profile photos can be linked to her identity with high possibility, as shown in Figure \ref{fig:linkedin}.

\vspace{2pt}\noindent \textit{Leak through file name:} We found several apps organize data related to a person or friends into a dedicated file named with sensitive or non-obvious sensitive information. For example, \texttt{WhatsApp} stores user's friends photos in the path \texttt{$\sim$/WhatsApp/Profile pictures/} with that friend's phone number as file name. They seem meaningless but could have significant privacy implications when combined with other public information, for example, the data from social networks. A sample attack will given in Section~\ref{sec:attack} to show that how these non-obvious sensitive information could be used in privacy attacks.

\vspace{2pt}\noindent \textit{Leak through folder name:} Some apps use account name as folder name directly, like \texttt{Renren}, \texttt{Momo}. \texttt{KaokaoTalk} will create a folder with the same name in two users' phones if they chatted with each other. We also find that a file sent to each other will have the same file name and saved in the same path. Such a naming convention reveals the connections among people, and even can be leveraged to infer user's chat history.

\vspace{2pt}\noindent \textit{Leak through other file:} Files left by uninstalled apps or put by the user may also disclose user's information, like email addresses. We leverage the command \textit{/system/bin/sh -c grep -r @xxx.com path} to match and extract email addresses from files (only three types, .txt, .doc(x), .pdf) in shared storage.

\vspace{8pt}
\noindent\textbf{What information has been leaked}. As shown in Table~\ref{privacycontent}, there is indeed some important sensitive information leaked through the app-private data on shared public storage. To better understand the privacy implication of such leak, it is better to use

Personal Identifiable Information (PII)~\cite{mccallister2010guide}, a well-known definition for private data, to classify and evaluate the leaked information. We defined two categories of sensitive data, \textit{Obvious sensitive data} and \textit{Non-obvious sensitive data}, by refining the concept of PII as below:

\begin{itemize}
\item \textbf{Obvious sensitive data}. It contains identifiers in PII related to user's real-world identity, including full name, phone numbers, addresses, date of birth, social security number, driver's license id, credit card numbers, etc.
\item \textbf{Non-obvious sensitive data}. It contains identifiers in PII related to user's virtual-world identity and also her friends' information. The virtual-world identifiers include email addresses, account name, profile photos, etc. We also consider the friends' information as shown in previous works~\cite{narayanan2009anonymizing}, they can be used to uniquely identify the user.
\end{itemize}

According to the definition, we divided the surveyed apps into three categories with privacy protection level from high to low, based on the type of sensitive data revealed. Table~\ref{privacyevaluation} shows the results of our evaluation.
\begin{table}[htb]
\centering
\caption{Privacy protection level of the 17 apps}
\newsavebox{\tableboxc}
\label{privacyevaluation}
\begin{lrbox}{\tableboxc}
\begin{tabular}{l|l|l} \hline

\textbf{Privacy level}&\textbf{App Name}&\textbf{Remarks}\\ \hline
 $\bigstar$$\bigstar$$\bigstar$     & \tabincell{l}{Facebook, Twitter,\\ Instagram, Skype} & No comments  \\ \hline
 $\bigstar$$\bigstar$     & \tabincell{l}{Line, Vine, WeChat} & \tabincell{l}{Audio files \\without encryption}  \\ \hline
 $\bigstar$ & \tabincell{l}{WhatsApp, Linkedin,\\ Viber, KakaoTalk, \\Tencent QQ, Alipay, \\Renren, Weibo,\\ Momo, EasyChat} & \tabincell{l}{Detailed problems are \\shown in Appendix~\ref{appendix2}} \\ \hline

\end{tabular}
\end{lrbox}
\scalebox{0.85}{\usebox{\tableboxc}}
\end{table}

\noindent\textbf{How to infer user's identity}. Without doubt, user's identity information is a key issue in any privacy research. We present 5 approaches to get a user's identity information by leveraging the sensitive data acquired, shown as follows:

\begin{itemize}

 \item A user's name and profile photo can be acquired from Viber.

 \item A user's identity can be acquired by searching her Renren UID.

 \item A user's identity can be acquired by searching her Weibo Username/UID.

 \item A user's identity can be acquired by searching the user's Linkedin profile photo on Google image search.

 \item We could find someone on Facebook with a high probability by the email addresses extracted from shared public storage and also the usernames acquired from other apps, since people prefer to use the same username and email address among their various social networking apps \cite{sameusername}.
\end{itemize}

Apparently, leaking obvious sensitive data should be prohibited and requires immediate actions from the app developers and Android developing team. The damage of leaking non-obvious sensitive data is less clear. For instance, attackers holding a user's photo still have no way to learn her real-world identity unless the user is famous or acquainted to attackers. However, our research shows it is possible to infer obvious sensitive data from non-obvious ones (in Section~\ref{sec:attack}) and the latter should also be well protected.

\subsection{Investigation on apps with a large scale}
\label{investigationlargescale}

The above analysis shows that all 17 popular apps have saved their app-private data to shared public storage, and 10 of them have leaked lots of privacy-related information. In order to have a better understanding of the scale and seriousness of this issue, it is necessary for us to investigate more apps. However, this is a difficult task in general, because accurate evaluation relies on intensive use of these apps, including account registration, sending message, and etc. Too many deep simulation of people daily use operations will make dynamic analysis tools not effective, especially can not scale to the large number of Android apps. Therefore, we take static analysis instead, the basic idea is scanning the app code for specific patterns that indicates the written of app-private data on shared public storage. Again, we only focus on the three categories that are most likely to touch user's sensitive information, so we selected 1648 different apps from our app repository (including 34369 apps downloaded previously).

The previous static analysis on apps, like \cite{gibler2012androidleaks, au2012pscout}, convert the app code from the DEX format to a JAR or Java source code, and then leverage WALA \cite{finktj} or Soot \cite{vallee1999soot} to complete the further analysis. Such analysis methods are intuitive and easily to operate, but in general, some information will be lost when converting app code to JAR or Java source code. Therefore, we leveraged Apktool~\cite{Apktool} to decompile apk to \texttt{smali} code which contain all the original apk information, and even can be repackaged to a new app. Then we did static analysis directly on the \texttt{smali} code.

%partly because apps include native code
Determining whether app-private data is really sensitive or not requires more knowledge of the app itself and even combining with other app's information, therefore is difficult to automate. We think an app is highly risk if it intends to create ``sensitive'' app-private folders or files on shared public storage. Specifically, ``sensitive'' app-private folder or file needs contain at least one of the following keywords, including \texttt{log}, \texttt{cache}, \texttt{files}, \texttt{file}, \texttt{data}, \texttt{temp}, \texttt{tmp}, \texttt{account}, \texttt{meta}, \texttt{uid}, \texttt{history}, which are learned from the apps by heuristic. For each keyword, we define our patterns similar to ``whole word match'' in Linux, but has more restrictions. For example, the patterns of keyword  \texttt{log} contain folders (\texttt{/user\_log}, \texttt{/log}, ...) and files (\texttt{user\_log.txt}, \texttt{user.log}, ...). What's more, we build a control flow graph (CFG) of an app's \texttt{smali} code to confirm whether the ``sensitive'' data is truly written on the shared public storage, since the data may be written in internal storage or some other secure places. In general, there are several ways to get the path of shared public storage, we summarize them and shown in Table \ref{pathexternal}. Also, we consider developers use general methods to create files (e,g., \texttt{FileOutputStream}) and folders (e.g., \texttt{mkdir}), and ignore the situations that developers use linux system commands to create files (like \texttt{touch}).

We marked each function $f$ in the CFG based on the following three criteria:

\begin{itemize}

 \item Whether it contains at least one of the methods described in Table \ref{pathexternal}.

 \item Whether it contains at least one of the patterns we defined above.

 \item Whether it uses \texttt{FileOutputStream} to create files or \texttt{mk-\\dir} to make folders.

\end{itemize}

\begin{table}[htb]
\centering
\caption{Methods of getting path of shared public storage}
\newsavebox{\tableboxz}
\label{pathexternal}
\begin{lrbox}{\tableboxz}
\begin{tabular}{l|l} \hline

\textbf{\tabincell{l}{Category}}& \textbf{\tabincell{c}{Methods}}\\ \hline

\multirow{6}{*}{Call API} & \texttt{getExternalStorageDirectory()} \\ \hhline{~-}
     & \texttt{getExternalStoragePublicDirectory()} \\ \hhline{~-}
     & \texttt{getExternalFilesDir()} \\ \hhline{~-}
     & \texttt{getExternalFilesDirs()}  \\ \hhline{~-}
     & \texttt{getExternalCacheDir()}  \\ \hhline{~-}
     & \texttt{getExternalCacheDirs()}  \\ \hline

Hardcode Path& ``\texttt{/sdcard}'', ``\texttt{/sdcard0}'', ``\texttt{/sdcard1}'' \\ \hline

\end{tabular}
\end{lrbox}
\scalebox{0.8}{\usebox{\tableboxz}}
\end{table}

We implemented our detecting method (shown in Algorithm 1) on the marked CFG after the above processes, the depth parameter was set as 3 according to our experience. The results show that 489 apps from the 1648 apps being analyzed intend to write some app-private data to shared storage. However, such a static scanning method has its limitations. So, we randomly chose 30 apps from the 489 suspicious apps, and manually operated these apps and did our evaluation, just like the work performed on the popular apps. The result shows that 27 apps truly wrote app-private data in shared public storage, which indicates such a privacy leakage problem revealed in this paper is widely exist and very serious among apps. Also, the false positive of our method is low.

\renewcommand{\algorithmicrequire}{\textbf{Input:}}
\renewcommand{\algorithmicensure}{\textbf{Output:}}

\begin{algorithm}[htb]
\label{alg:algorithm1}
\caption{: Detecting vulnerable apps}
\begin{algorithmic}[1]
\Require Class\_set $C$, Keyword\_Patterns\_set $KS$, Path\_API\_set $PA$, Write\_API\_set $WA$
\Ensure bool sensitive

\For{class c in $C$}
\For{function $f$ in c}
\State condition.clear();
\State DFS($f$, depth);
\If{condition == Union($KS$, $PA$, $WA$)}
\State return true;
\EndIf
\EndFor
\EndFor
\State return false;

\\

\Procedure {DFS}{function $f$, int depth}
\If{depth == 0}
\State return;
\EndIf
\For{all element e in $f$.mark}
\State condition(e)=true;
\EndFor
\For{all callee ce of $f$}
\State DFS(ce, depth-1);
\EndFor
\EndProcedure

\end{algorithmic}
\end{algorithm}

\section{Attacks based on non-obvious \\sensitive data}
\label{sec:attack}

In this section, we present three example attacks based on the non-obvious sensitive information extracted from app-private data saved on shared public storage. We begin with a brief introduction to the design of a malicious Android app called SAPD (``Smuggle App-Private Data''), followed by detailed description of three concrete attacks: attacking victim's location, identifying the owners of the phone numbers acquired from WhatsApp friends list, and even getting chatting buddy and user's chat history. Again, we want to emphasize is that the three attacks are different from previous attacks which have been reported.

\subsection{Attack preparation}

Since it is assumed that attackers have already studied the vulnerable apps and know where and how to get useful information from app-private data, the natural next step is to develop a malicious app that could dig out and upload such information to a server under their control. As the result, the malicious app need have permissions to access (more accurately, to read) shared public storage and Internet. As discussed in section~\ref{adversarymodel}, such permissions are common and not alarming.

The weakest part of this malicious app might be the potential outstanding network traffic footprint, especially for users with limited 3G plan. We implemented two optimizations in our app prototype SAPD to get around this limitation. First, try to minimize the uploaded data. For example, for information leaked through file name or folder name, it is only necessary to upload the list of file names instead of the whole file. Another optimization is to upload data only when a Wi-Fi network is available. It can be achieved by simplely using the \texttt{WiFiManager} class provided by Android SDK, with the cost of requesting an extra permission \texttt{ACCESS\_WIFI\_STATE}. However, SAPD achieved the same goal by reading public available files in \textit{procfs} from Linux kernel of Android. More specifically, Android put the parameters of Address Resolution Protocol in file \texttt{/proc/net/arp} and other wireless activities in \texttt{/proc/net/wireless}, so by reading above files, SAPD is able to know whether a WiFi network is connected or not.

After having extracted the useful information and identified appropriate data uploading timing, the last thing that SAPD will do is just sending the data out to a malicious server where all subsequent attacking steps will taken place. To minimize the possibilities of being caught due to suspicious CPU usage or abnormal battery consumptions, SAPD only reads and uploads filtered useful data, and it will never perform any kind of intensive computations or Internet communications.

\subsection{Inferring your location}
\label{sec:attack3}

\begin{figure*}[ht]
    \centering{
		\includegraphics[scale=0.35]{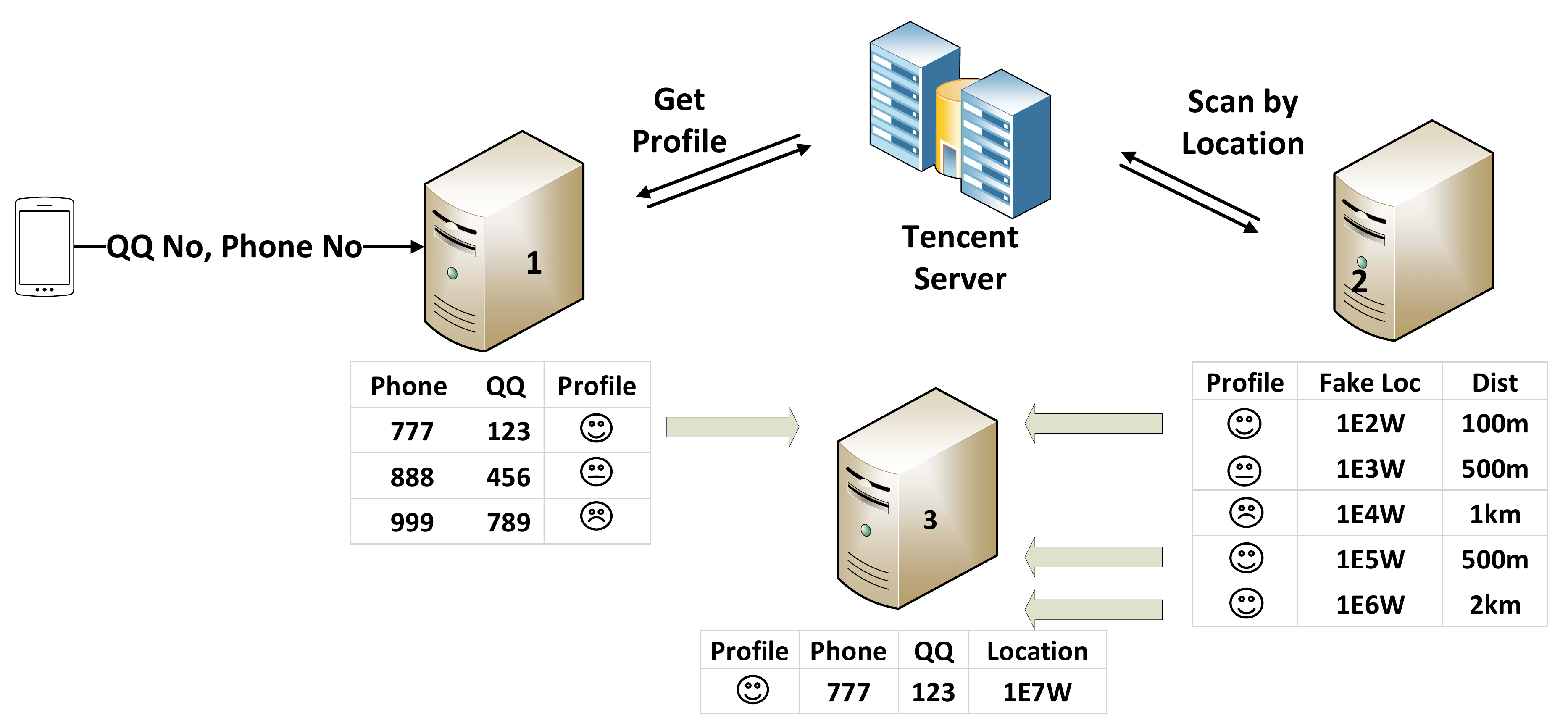}
		\caption{The schematic diagram of location attack}
		\label{fig:location1}}
\end{figure*}

In this section, we describe our attack on inferring user's location from data in shared storage. Location of a phone user is considered as sensitive from the very beginning and there are already a lot of research works on inference attacks and protections. In recent years, location-based social discovery (LBSD) is becoming popular and is widely adopted by mobile apps. The locations of users are updated and viewable by friends or even strangers. For the latter case, to protect user's privacy, only distance between the user and the viewer is revealed. However, recent works show such protection is very weak. Chen et al. analyzes LBSD in large-scale and also shows it is possible to re-identify user with high precision~\cite{chen2013and}. A recent work by Li et al. demonstrates that the geo-location of user can be revealed~\cite{li2013all}. Our attack also aims to identify user's location from LBSD network but we make improvements by combining the profile information extracted from victim's phone. We show the attack would be more accurate and could be a realistic threat. As a showcase, we demonstrate our attack on \texttt{WeChat} app.

\begin{figure}[t]
    \centering{
		\includegraphics[scale=0.46]{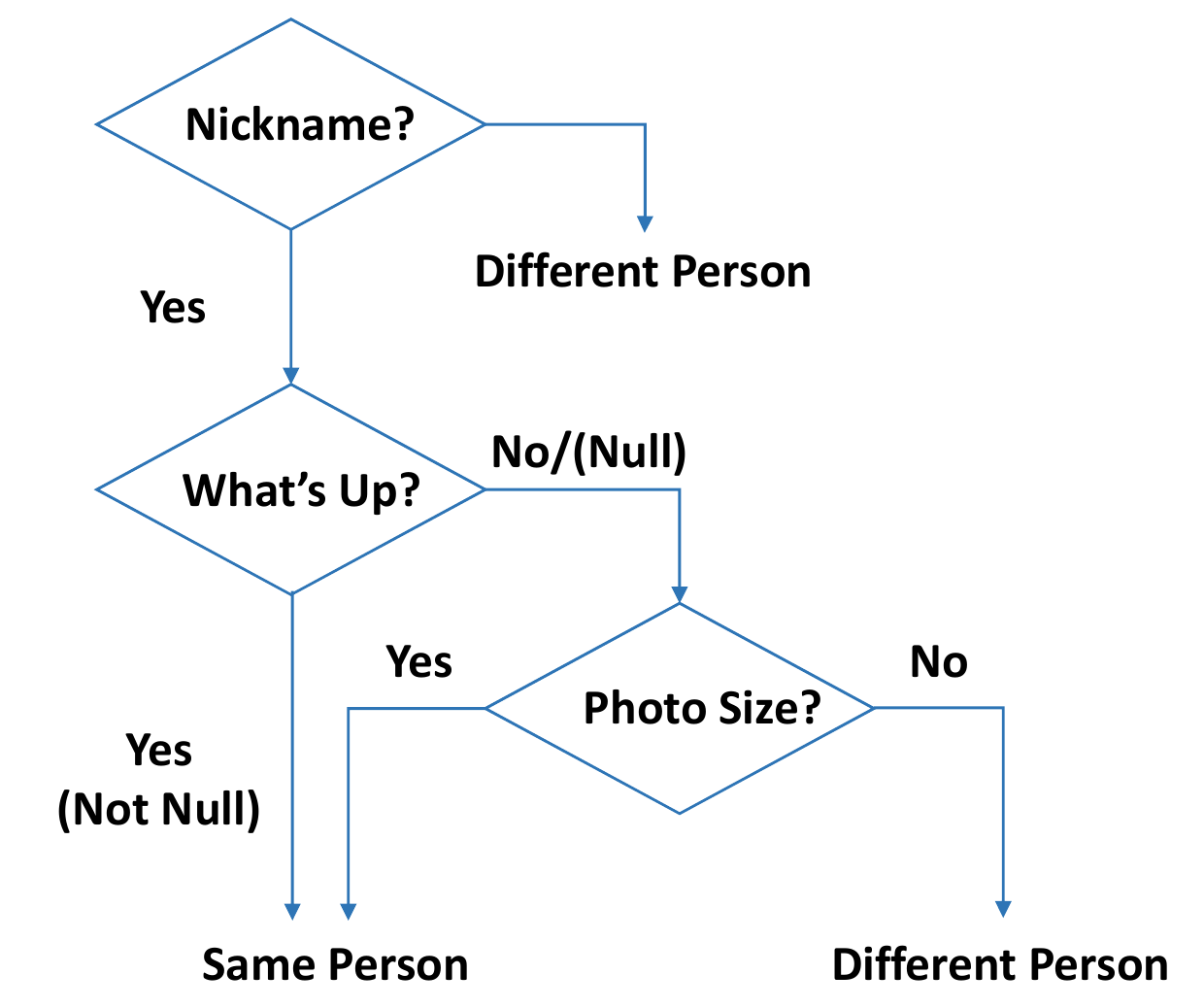}
	\caption{The processes of profile information comparison}
		\label{fig:compare1}}
\end{figure}

The LBSD module in \texttt{WeChat} is called ``People Nearby'', through which, the user can view information of other users within a certain distance. The public information includes nick name, profile photo, posts (called \textit{What's Up}), region (city-level) and gender. As described in section~\ref{findingpopular}, the user's phone number or QQ userid is left by \texttt{WeChat} on shared storage. These information will be collected by SAPD we built and sent to one of our servers (denoted as $S_1$). These servers are installed with emulated Android environment for running WeChat app. $S_1$ will query the server of Tencent (the company operating WeChat) for profile information. In the meantime, attacker needs to instruct another server (denoted as $S_2$) to run WeChat using fake geolocations, checks People Nearby and downloads all the profile information and the corresponding distances from discovered nearby users. The profile information from $S_1$ is then compared with the grabbed profile information (downloads from $S_2$) in another server (denoted as $S_3$) followed the steps shown in Figure~\ref{fig:compare1}. If a match happens,  $S_2$ will continue to query People Nearby for two more times using different geolocation (faked) to get two new distances. Through the three point positioning method, the target's accurate location can be inferred. Since the target user may keep moving, the three servers need to synchronize on user's information. Figure \ref{fig:location1} shows the processes of our attack and we elaborate the technical details in each stage as below:

\vspace{2pt}\noindent\textbf{Getting Users' WeChat Account}.
Though \texttt{WeChat} userid is not stored on shared storage, QQ userid and phone number are stored instead. They are binded to WeChat account and has to be unique for each user. Therefore, our app collects these two data and sends them to our server $S_1$ for profile querying.
\vspace{2pt}\noindent\textbf{Getting Users' Profile Information}.
Next, the attacker uses QQ userid or phone number to query Tencent server for user's profile information. The returned profile consists of 5 fields: nick name, profile photo, posts (\textit{What's Up}), region and gender. Our task is to assign the location information for this profile. Unfortunately, this field is not provided directly from Tencent server and updated according to the user's geo-location. What we do here is to run WeChat on another server $S_2$, frequently refresh People Nearby to get as many profiles and distances till a profile match happens, and use the corresponding distances to infer her location. A challenge here is to extract the profile and distances from WeChat app, as there is no interface exposed from WeChat to export these information. After we decompile its code, we found the app invokes an Android API \texttt{setText} from \texttt{android.widget.TextView} to render the text on screen whenever a profile is viewed. We therefore instrument this API and dump all the texts related to profiles into logs. This helps us to extract the \textit{Nickname, What's Up, and Region}. In general, we can distinguish a people from others based on these three features. However, it is also possible (with very low possibility) that different profiles have the same values of these three fields. If such a situation happens, we extract the size of profile photo and consider it in the matching process. We do not use the photo content since the comparing is more time-consuming and size is enough.

\vspace{2pt}\noindent\textbf{Comparing process}.
The comparison processes in $S_3$ are shown in Figure~\ref{fig:compare1}. The first step we need to compare the \textit{Nickname} from $S_1$ and $S_2$, if they are the same, we will continue to compare their \textit{What's Up} information, if the \textit{What's Up} are not blank and the same, we think the two users are the same person. If the \textit{What's Up} information is different or blank, we will continue to compare their profile photos' size for another round of check. We add this additional step since the user could post new \textit{What's Up} information while our profile information stored on $S_1$ has not been updated to the latest status. Here we use ``Shark for Root'' \cite{wireshark}, a network traffic capturer to intercept the packets and extract the photo size information from \texttt{pcap} file.

If a match happens, we get another two distances between the target and our faked coordinates to calculate the target's location. We use an app called ``Fake GPS location'' \cite{fakegps} to fake the server's geolocation to different places. To make WeChat work more stable when this app fakes GPS, ``Allow mock locations'' option has to be turned off. This is achieved through setting the access right of the directory \texttt{/system} as \texttt{rw} and then moving ``com.lexa.fakegps-1.apk'' from \texttt{/data/app} to \texttt{/system/app}. We also enable the ``GPS satellites'' option in Location services to let apps use GPS to pinpoint the location. Meanwhile, the ``Google's location service'' option was disabled to limit apps using data from Wi-Fi and mobile networks determining the approximate location.

We first set the default location of ``Fake GPS location'' to a spot within our university. For the densely populated places, we added several more anchor points. Since People Nearby only display limited amount of users (about 100), using more anchor points and merging the list of discovered people will increase the chance of hitting the target user. Next, we use a script to automatically refresh People Nearby. For each point, to load all the people's information appeared in People Nearby, the script clicks on the links to people's profile one-by-one through triggering event \texttt{KEYCODE\_DPAD\_DOWN}. This process has to request data from Tencent. To avoid raising alarm from Tencent, the script sleeps a while before changing to a new anchor point. If someone we have known appeared in the list of People Nearby, another WeChat will be launched to get new two distances by mocking its GPS to another two locations near the place. With the help of a program which implements the three point positioning algorithm, we can easily calculate her location.

\vspace{2pt}\noindent\textbf{Attack evaluation}.
We evaluate our attack on 20 participants. Each participant has installed WeChat with People Nearby turned on (so their profiles will be open to view). \ignore{Should have concrete result...} Our attack successfully reveal the live locations for most of the participants and are verified by them. To point out, some of the inferred locations are not exact where the user stays, but they are all within accepted within the acceptable range.

\begin{figure}[ht]
    \centering{
		\includegraphics[scale=0.8]{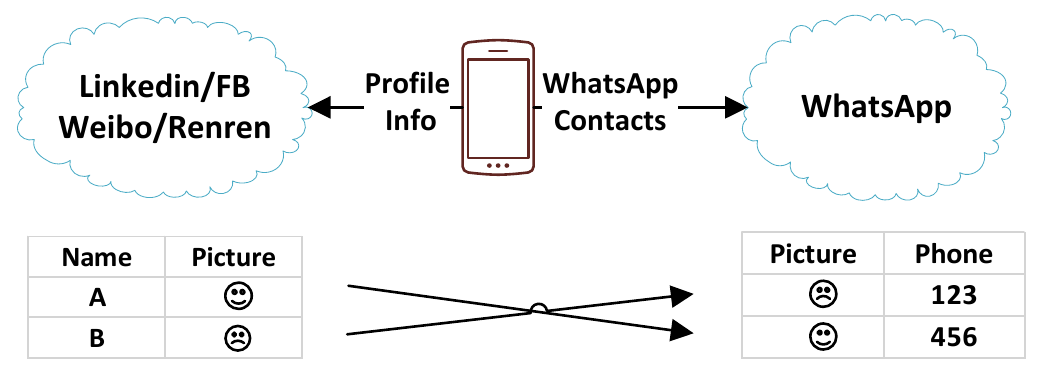}\\
	\caption{The diagram of attack user's WhatsApp friends}
		\label{fig:phonenumber}}
\end{figure}

\subsection{Inferring your friends' identities}
\label{sec:attack2}

Besides inferring the identity of the phone owner, the identities of her friends are also attractive to the adversary, which could lead to more accurate profiling of the phone owner or attacks against her friends. However, achieving this goal for attacker is not straightforward. Let us take a look at \texttt{WhatsApp}, only the profile photos and phone numbers of the user's friends are accessible and it is difficult for adversary to directly learn who they are. On the other hand, studies have shown the internet users tend to use default privacy settings of social network and expose their information (including full name) to visitors even strangers~\cite{SocialMediaPrivacy}. Moreover, people tend to use the same photo for different social networks~\cite{SameProfilePicture}. These two observations inspire us to match the profile photos between \texttt{WhatsApp} and social networks to infer the identities. Figure \ref{fig:phonenumber} illustrates this attack and we elaborate the steps below:

\vspace{2pt}\noindent\textbf{Getting photos from WhatsApp.}
After adversary obtains the photos in victim's phone, she could opt to transmit all of them to the server for later analysis. However, this will incur a lot of network overhead and is less stealthy. Instead, we leverage a special feature of \texttt{WhatsApp} server on adding friends to reduce the volume of data required for transmission. Specifically, a registered \texttt{WhatsApp} user (say $U_A$) can get the profile photo of any other user (say $U_B$) by adding phone number of $U_B$ into $U_A$ contact list, and there is no need for permission from $U_B$. Therefore, our malicious app \texttt{SAPD} collect the phone numbers and compile them into a standard contact file (with file extension \texttt{.vcf}) and send to our server. After that, the attacker loads the file into her smartphone's contact list and then launch \texttt{WhatsApp} to download the profile photos. To notice, downloading a profile photo is only triggered when the account owner clicks user' profile photo. Therefore, we write a monkeyrunner script to automate this process by simulating the clicks. Comparing to directly sending photos, this approach can reduce about 3MB network traffic (taking 50 photos as an example) and hence is very helpful to make SAPD more stealthy.

\vspace{2pt}\noindent\textbf{Getting photos from social networks.}
Since an internet user is likely to use default privacy settings on social networks, her user name, full name (if required), friends and etc. are usually open to public. In particular, her friends' photos are usually grouped together and stored into a her public gallery (e.g., user X's friends' photos are stored under \url{https://www.facebook.com/X/friends}). We validate this setting in popular social network sites, including Facebook, Linkedin, RenRen and Weibo and we believe this setting should apply in most cases. In order to grab these photos, user's corresponding social network userid is required and is usually not direct available for the adversary. However, a large number of users also install the mobile clients of these social networks and our study already shows they tend to leave userid in shared storage within the reach of attackers (Linkedin, RenRen and Weibo). In case the userid is not directly revealed, we found user's email can be used in replace (Facebook). As shown in Section~\ref{findingpopular}, emails can be grabbed through scanning shared storage. After retrieving user's userid or email, we use a crawler built by ourselves to query the websites and download all the friends' photo and profile information (e.g., full name and address) into attacker's server.

\vspace{2pt}\noindent\textbf{Comparing photos.}
After we have the photos from \texttt{WhatsApp} and social networks, the next step naturally is to find matches. There are already a bunch of image matching techniques proposed and some advanced ones are able to recognize faces and match the photos of one person in different context. Using the advanced ones would improve the chance of successful matching but it is also time-consuming. In fact, simple algorithm, like sampling regions and calculating similarity, is enough for our scenario as user prefer to use the same profile photo in different social networks. Therefore, we adopt a simple matching algorithm described in~\cite{picturecompare1}.

\vspace{2pt}\noindent\textbf{Attack evaluation.}
We evaluate this attack by running our developed app on the participant's phone and also collect her friends' profiles using the server we set up. The app is able to collect 50 profiles containing phone numbers and photo from her \texttt{WhatsApp} contact. By searching the photos, we collect 29 social network profiles (12 from Linkedin, 19 from Facebook, 11 from Renren and 5 from Weibo, some are the same people) and they are all correctly matched as verified by the participant. This result indicates the phone user's friends' identities could be inferred with high chance through our attack.

\begin{figure}[ht]
    \centering{
		\includegraphics[scale=0.18]{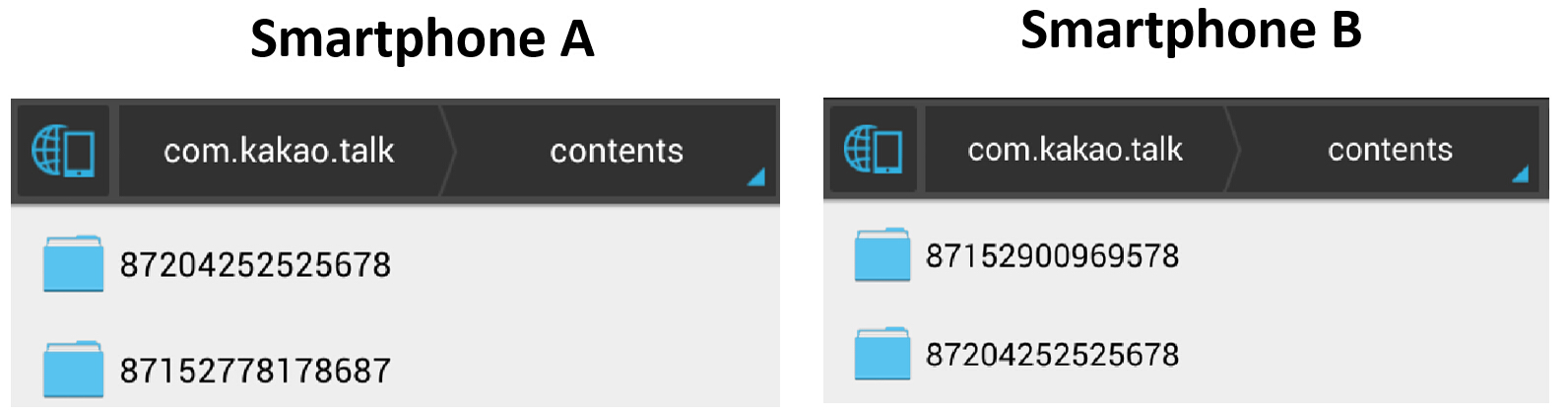}
	\caption{The contents folders of KakaoTalk on two smartphones}
		\label{fig:kakaotalk}}
\end{figure}

\subsection{Inferring your chatting buddy}
\label{sec:connection_attack}

Instant messaging apps usually keep the chatting logs under shared storage. In some cases, the logs are stored in plain text (e.g., \texttt{Easy\\Chat}) and the list of user's chatting buddies will be directly exposed to the adversary. Encrypting the logs seems to fix the problem, but our attack on ~\texttt{KakaoTalk} shows the problem is still not solved.

In fact, ~\texttt{KakaoTalk} creates an tag when conversation happens and then dumps the attachments (e.g. photos and audio files) under a folder named using the tag. As shown in Figure \ref{fig:kakaotalk}, the tag between smartphone A and smartphone B is \texttt{87204252525678} (a 14-digit number). Though it is difficult to recover user IDs from the tag, it is still feasible to infer the chatting buddies.

Assuming the malicious app SAPD is installed on a number of phones, it will keep collecting the tags and transmitting them to attacker's server. When tags from two different phone match, it is clear that the owners of the phones have chatted some time. The chance of matching is relatively small considering the large user base of~\texttt{KakaoTalk}, however, attacker can leverage social-engineering tricks (e.g., sending spams) to propagate SAPD from the infected user only to her friends to increase the success rate. After the entities of conversation inferred, the attacker could further recover the time point for each chat by leveraging the creation time of attachment files.

\section{Defense}
\label{sec:defense}

Our survey on popular Android apps shows leaks of sensitive information from app-private data stored on public storage do exist. Given the number of all apps in the wild, the issues we found are very likely to be a tip of the iceberg. Against this problem, several guidances and solutions are proposed but all of them incur huge efforts from app developers and users. Instead, we propose a new framework which extends the existing Android system and truly isolates app-private data from different apps.

\subsection{Existing solutions}
Since there is no finer-grained protection upon public storage, the data needs be carefully examined before saving to it. Presently, there is no tool as far as we know capable of automating this checking process, and the developers have to manually scrutinize their code and keep sensitive data protected. The CERT Oracle Coding Standard for Java~\cite{long2011cert} states two coding guidances regarding storing sensitive information:
\begin{itemize}
\item \textit{Encrypting data}. Before an application saves sensitive information to public storage, it needs to be encrypted. Following this rule would mitigate most data leakage problems (information may still be leaked through side-channels, like file size), but unfortunately this rule is not correctly enforced in practice. Though Android provides cryptographic API, many developers make critical mistakes on using them, including using ECB mode and non-random IV for CBC encryption~\cite{egele2013empirical}, and sensitive information from the encrypted files could be partially or even fully revealed to attackers. Ironically, even the developers of most popular apps make critical mistakes and directly reveal the encryption keys~\cite{whatsappuserchat}. What is worse, our research shows even developers correctly use encryption routines, they do not always identify the sensitive objects. As shown in Section~\ref{findingpopular}, user's account number and phone number e.g., are used as filenames, which can be abused to infer a wide spectrum of users' sensitive information.
\item \textit{Saving to internal storage}. Android uses Ext2/4 for internal storage and an app can only access its own data from internal storage. Therefore, it is encouraged to keep sensitive data there. Specifically, data can be saved to a file created in application data directory with permission set to \texttt{MODE\_PRIVATE}. However, developers would like to only decide very sensitive data and throw all the other data to public storage, such a habit mainly caused by history reasons which are elaborated in Section \ref{sec:discussion}.
\end{itemize}

Expecting all the developers to obey the security rules or fix security issues is impractical. On the other hand, users could protect themselves: by converting the format of public storage from FAT to Ext2/4, the built-in access control mechanism of Ext2/4 will be inherited and storage of different apps will be well separated. Yet, this migration might be only suitable for advanced users: the Android system has to be rooted and additional apps have to be installed to support Ext2/4 format public storage; Windows users have to install additional software to browse and read files on it; the existing files have to be backed up and restored when formatting. None of these steps are easy for common users.

\begin{figure}[t]
    \centering{
		\includegraphics[width=0.45\textwidth]{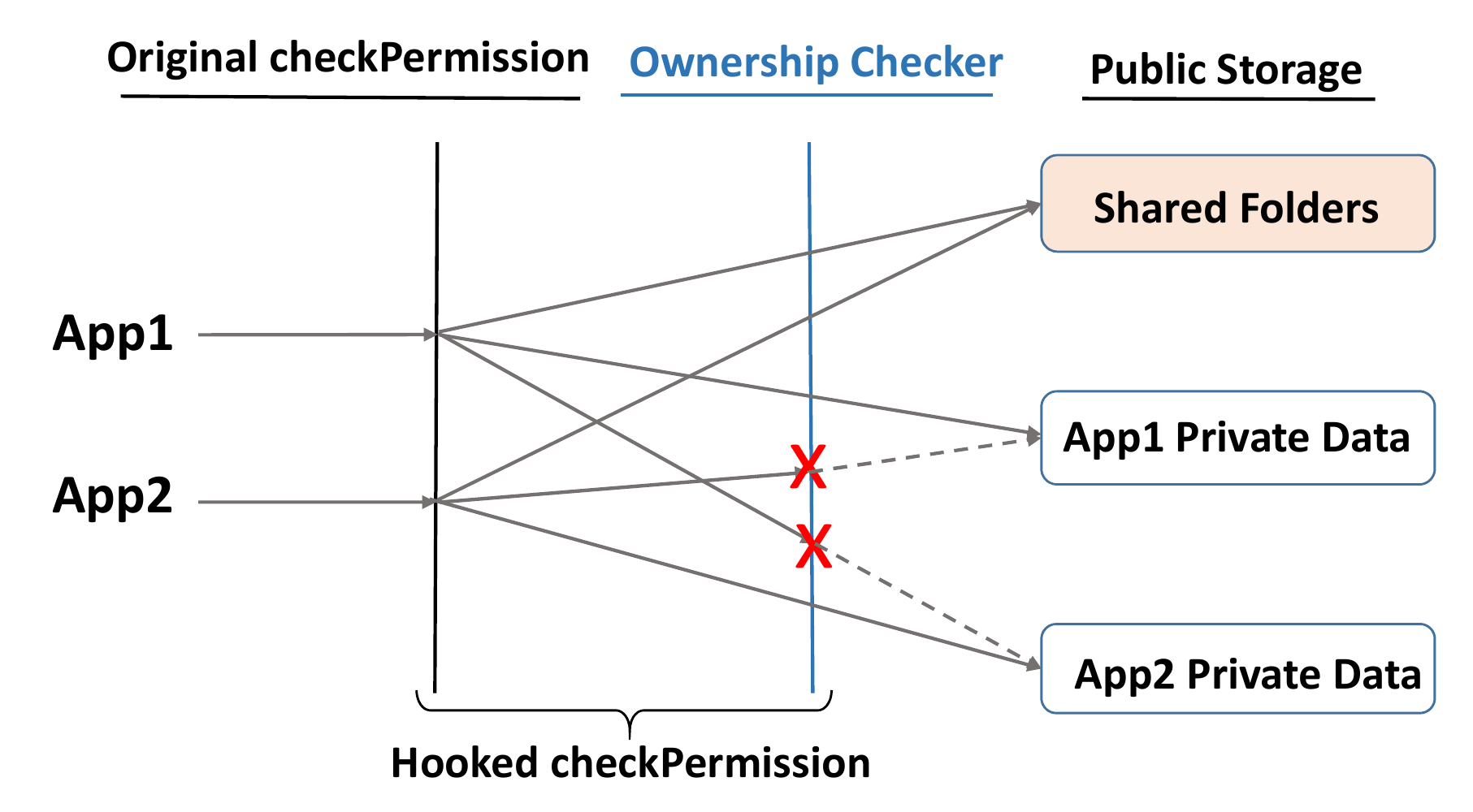}
    \caption{Proposed defense framework, adding ownership checker to restrict apps accessing other apps' private data.}
		\label{fig:defense}}
\end{figure}

\subsection{Proposed framework}
The problem will not be fixed in the near future if following these paths described above. On the contrary, modifying the Android system and push the upgrades to users' devices would be an easier way to mitigate the security issues. For this purpose, we propose to augment the existing security framework on public storage by instrumenting the API \texttt{checkPermission()} \cite{checkPermission}, the framework is described as below:

\noindent\textbf{Architecture}. We design a new module named \textit{ownership checker} as shown in Figure~\ref{fig:defense}, which works on Android Middleware layer and can achieve mandatory access control (MAC) for app-private data. Specifically, when the targets are public resources, like shared music directory, the access is permitted (certainly, the app needs to hold necessary permissions), while the targets are located in app's private folder, the access is only permitted when the calling app matches the owner, if not, the  \textit{ownership checker} will return PERMISSION\_DENIED even if the app has requested the \texttt{READ} or \texttt{WRITE} permission. To enforce such rule, we create a system file owner\_checker.xml storing the mapping between apps and resources, similar to Access Control Lists (ACL) of Ext4 file system. And the Android system code, which manages the \texttt{checkPermission()}, are modified to read the mapping before actual file operations. An exception will be thrown to the caller if a mismatch happens.

\noindent\textbf{Ownership inference}. The ownership mapping between apps and resources needs to be established before actual access happens, which, however, is not a trivial task. We have to deal with the situation that the public storage has already kept apps' data before our module is installed, while the owner of data is not tracked. To fix the missing links, we exploit a caveat regarding naming convention: an app usually saves data to a folder whose name is similar to its package name. For example, Viber saves data to \texttt{$\sim$/viber/} and its package name is \texttt{com.viber.voip} (app's UID and package name can be acquired from packages.xml under \texttt{/data/system}). Therefore, we initialize the mappings by scanning all the resources. For a given resource, we assign the owner app if the resource location and app package name share a non-trivial portion. To notice, this initialization step could not infer all the ownerships when an app stores the data in a folder whose name is irrelevant. While the access will be blocked for this case, we provide an interface for users to add the ownerships. When a failed access happens or the resource has no ownership associated, a dialog is prompted to user to ask if the access is permitted. If the permission is granted, a new mapping will be added. To reduce the hassles to users, the User-driven access control model~\cite{roesner2012user} can be integrated to automatically assign ownership based on user's implicit actions.

\section{Discussion}
\label{sec:discussion}

\vspace{2pt}\noindent\textbf{Evolvement of Android storage model}. In the early stage, Android phone is equipped with limited onboard Flash memory (only 256MB for the first android phone, HTC Dream). In the meantime, its storage can be expanded through large volume SD card (could be several GB) which is usually shipped together. This storage model forces Android app developers to save the most of the data to FAT formatted SD card, while the sensitive information, like passwords, is saved to more secured Ext2/4 formatted Flash memory. The size of onboard Flash memory is gradually enlarged, however, the app developers still prefer to use SD card, even forcing Android to simulate FAT format on a compartment of Flash memory. While keeping compatibilty for existing apps reduce the workload for developers, it brings severe security and privacy risks, as shown in our attacks. Notably, there have been continuous efforts on improving the security of storage model by Google, including introduing \texttt{READ\_EXTERNAL\_STORAGE} permission in Android 4.1 and \texttt{WRITE\_MEDIA\_STORAGE} in Android 4.4. Unfortunately, these enhancements do not solve the fundamental issues with the storage model. We proposed a new framework which enforces finer-grained access control mechansims on the insecured part of storage system (SD card and part of Flash memory simulating SD card). This approach does not require changes from existing apps or intensive efforts from users and could be easily applied.

\vspace{2pt}\noindent\textbf{Limitations of app study}. We launched a study on existing apps to understand the scale of the problem (see Section~\ref{investigationlargescale}). We built a tool running static analysis on app's code and use a set of heuristics to determine if the app saves sensitive information to unprotected shared storage. This simple tool identifies a large number of potentially vulnerable apps and show reasonable accuracy from our sampling result. However, it is inevitably suffers from false negatives (e.g., the file name does not contain the keywords we used) and false positives (e.g., the information saved is not sensitive). We leave the task of building a more accurate detector as future work.

\section{Related Work}
\label{sec:relatedWork}

\noindent\textbf{Harvesting users' sensitive data from mobile devices}. Since mobile devices now hold a lot of people's private information, they have become popular targets for attackers. Attacks like stealing users' chat history~\cite{whatsappuserchat} have been proved their feasibility in the real world. These attacks usually depend on certain vulnerabilities identified from the victim apps. Instead, our attacks exploit a general design flaw of Android system - the lack of finer-grained control on shared storage, and we demonstrated the effectiveness of attacks on a number of popular apps from different categories.

In addition to steal user's sensitive information directly, a lot of research focused on inferring user's private information, especially the location. Srivatsa et al. showed a set of users' location traces can be de-anonymized through correlating their contacts with their social network friends' list~\cite{srivatsa2012deanonymizing}.  Using users' public comments, Zheng et al. showed it is feasible to infer users' locations and activities ~\cite{zheng2010collaborative}.  Also a recent work~\cite{zhou2013identity} by Zhou et al. targets to infer information of users from more perspectives, including identities, locations and health information. While their works exploit side channels of Android system, our work takes advantage of the data on shared storage and corresponding attacks are easier to carry out.

\noindent\textbf{Inferring users' information from social network}. Social networks allow users to share information with their friend. It is another hub of user sensitive information and its privacy issues have been well studied in several previous papers~\cite{mislove2010you, wondracek2010practical, balduzzi2010abusing, goga2013exploiting}. In~\cite{mislove2010you}, Mislove et al. found that users with common attributes are more likely to be friends and often form dense communities, and they proposed a method of inferring user attributes that is inspired by previous approaches to detecting communities in social networks. Goga et al. proposed an attack to link the user's identity to her accounts on different social networks~\cite{goga2013exploiting}. This attack exploits only innocuous activity that inherently comes with posted content. Also user's group memberships~\cite{wondracek2010practical} and the ``friend finder'' function~\cite{balduzzi2010abusing} also could leak user's identity (de-anonymization attack).  They studied how to use the network characteristics to infer the social situations of users in an online society.

These works focused on social networks themselves, that is to say, all analyzed data come from one or several social networks. However our attack combines the sensitive data from users' phones (shared storage) and public data from social networks, which is more powerful.

\noindent\textbf{Protecting users' privacy on mobile devices}. To defend against the existing or potential attacks tampering user's privacy, a bunch of defense mechanisms have been proposed. Roesner et al. proposed user-driven access control to manage the access to private resources while minimizing user's actions~\cite{roesner2012user}. Ongtang et al. proposed a finer-grained access control model (named Saint) over installed apps~\cite{ongtang2012semantically}. It addressed some limitations of Android security through install-time permission-granting policies and runtime interapplication communication policies. Mobiflage~\cite{skillen2013implementing}, developed by Adam Skillen et al., was leveraged to enable Plausibly Deniable Encryption (PDE) on mobile devices by hiding encrypted volumes within random data on a device external storage. Users are needed to comply with certain requirements, but Mobiflage also has highly risks to be weakened even if users follow all the guidelines which makes it not practical. To protect users' location, several methods were proposed, like~\cite{shokri2012protecting} which restrict the information disclosed from locations.

Besides, efforts have also been paid on code analysis to block the information leakage. Enck et al. developed TaintDroid~\cite{enck2010taintdroid} to prevent users' private data from being abused by third party apps. DroidScope~\cite{yan2012droidscope} is another dynamic Android malware analysis platform. Based on the idea of virtualization, it reconstructed both the OS-level and Java-level semantics simultaneously and seamlessly.

\section{Conclusion}
\label{sec:conclusion}

It is known that the shared public storage on Android is insecure, due to its coarse-grained access model. Therefore, it is highly recommended that the sensitive data should be avoided from saving there. In this paper, we carry out a large-scale study on exsting apps on whether app developers follow this rule and the result turns out to be glooming: a siginificant number of apps save sensitive data into the insecure storage, some of the apps are even top ranked in Android market. By exploiting these leaked data, it is possible to infer a lot of information about the users, severely violating users' privacy. The security of users' data under current model heavily rely upon app developers, but requiring all app developers identify the sensitive objects and correctly handle it is impractical. Instead, we propose a new defense framework by extending the existing Android systems which achieve both compatibilty and finer-grained access control. To summarize, our study reveals unneglectable security issues in shared storage and further actions need to be taken to fix the problems.

\bibliographystyle{abbrv}
\bibliography{main}
\appendix
\section{The versions of the 17 apps installed on our smartphones}
\label{appendix1}

\begin{table}[htb]
\centering
%\caption{The versions of the 17 apps installed on our smartphones}
\newsavebox{\tableboxa}
\label{applist}
\begin{lrbox}{\tableboxa}
\begin{tabular}{l|l} \hline

\textbf{App Name} &\textbf{Version installed}\\ \hline

Facebook &\texttt{13.0.0.13.14}  \\ \hline
Instagram&\texttt{6.2.2}\\ \hline
Twitter  &\texttt{5.18.1}\\ \hline
Linkedin &\texttt{3.3.5}\\ \hline
Vine &\texttt{2.1.0}\\ \hline
Weibo &\texttt{4.4.1}  \\ \hline
Renren & \texttt{7.3.0}\\ \hline
Momo& \texttt{4.9} \\ \hline
WhatsApp &\texttt{2.11.186}\\ \hline
Viber &\texttt{4.3.3.67} \\ \hline
Skype &\texttt{4.9.0.45564}\\ \hline
Line & \texttt{4.5.4} \\ \hline
KakaoTalk &\texttt{4.5.3}\\ \hline
Tencent QQ & \texttt{4.7.0}\\ \hline
WeChat& \texttt{5.2.1}\\ \hline
EasyChat& \texttt{2.1.0}\\ \hline
Alipay &\texttt{8.0.3.0320}\\ \hline

\end{tabular}
\end{lrbox}
\scalebox{0.85}{\usebox{\tableboxa}}
\end{table}

\section{The details of user privacy data extraction }
\label{appendix2}

\noindent {\bf Viber} is an instant messaging app with 300 Million users \cite{10MessagingApps}. The text file \texttt{.userdata} saved under \texttt{$\sim$/viber/.viber/} reveals a lot of user's information, including the user's real name and phone number. Besides, it also contains the path pointing to the user's profile photo, e.g. \texttt{$\sim$/viber/media/User Photos/xxx.jpg}.

\vspace{2pt}\noindent {\bf WhatsApp} is an instant messaging app with 450Million users \cite{10MessagingApps}. The user's profile photo is stored under the directory \texttt{$\sim$/WhatsApp\\/.shared/}, with file name \texttt{tmpt}. The profile photos of user's friends are saved under the directory \texttt{$\sim$/WhatsApp/Profile pictures/} after the profiles are viewed, and they are named by profile owners' phone numbers without any obfuscation.

%It caches the photos viewed and downloaded to the phone, including both user's profile photo and her friends' photos.

\vspace{2pt}\noindent {\bf Linkedin} is a business-oriented social networking app with 300 Million users \cite{Linkedinuser}.  This app cache the photos into the directory \texttt{$\sim$/Android/data/com.Linkedin.android/cache\\/li\_images/}.  Though the images stored under the directory can be many, the user's profile photo is unique in file size and modified time. In general, the user's profile photo is about 30KB, while others are about 3KB (if the user does not see the HD profile photo).

\vspace{2pt}\noindent {\bf KakaoTalk} is an instant messaging app with 100 Million users \cite{10MessagingApps}.  If user A chats with user B, the app will create a content folder has the same name in both the Android phones, under the path \texttt{$\sim$/Android/data/com.kakao.talk/contents/}. ~The\\ files on the two phones also have the same name (10 digital numbers), size and stored in the same path, like \texttt{$\sim$/Android/data/com.\\kakao.talk/contents/14 digital numbers/xx/}.

\vspace{2pt}\noindent {\bf Tencent QQ} is an instant messaging app in China with 816 Million users \cite{allusers}.  Under the path \texttt{$\sim$/tencent/com/tencent/mobi-\\leqq/}, it is very easy for us to acquire a user's  QQ account from the log file named as ``com.tencent.mobileqq.xxx.log''.

\vspace{2pt}\noindent {\bf Weibo} is a Chinese microblogging service with 500 Million users \cite{Weibouser}. There is a file named by the user's uid under the path \texttt{$\sim$/sina/weibo/page}, and we can acquire the user's username and even her email address if she used it to register her account.

\vspace{2pt}\noindent {\bf Alipay} is a online payment service in China with 300 Million users \cite{Alipayusers}. We could find a user's phone number from two cache files, one is the \texttt{\_meta} file in the directory \texttt{$\sim$/com.eg.android.Alipa-\\yGphone/cache/}, the file content reveals the user's phone number directly, and also points out the other file in the same directory which discloses the user's phone number.

\vspace{2pt}\noindent {\bf Renren} is a social networking app with 194 Million users \cite{Renrenusers}. A folder named by the user' UID is stored in the directory \texttt{$\sim$/Andro-\\id/data/com.renren.mobile.android/audio/cache\\/JasonFileCache}. Even user' visit histories are also stored in this folder, which contains the name, UID of user's friends. The audio files are named as the format \texttt{UID+hash value}, which also disclose the user's UID. We can find the user's personal home page by the URL \textit{http://www.renren.com/UID} in a browser.

\vspace{2pt}\noindent {\bf Momo} is a location-based services instant messaging app with 100 Million users \cite{Momousers}. We can find a folder named as the user's account in the directory \texttt{$\sim$/immomo/users/}. Through searching the account number, we can not only get her Momo profile information, but also infer her location and even track her.

\vspace{2pt}\noindent {\bf EasyChat} is an instant messaging app with 60 Million users \cite{Yixinusers}. The file \texttt{pjsip\_log.txt} under the directory \texttt{$\sim$/Yixin/log/} shows us all the call records information in plaintext. For example, we can extract information ``From: 8525984xxxx, To: 1471434xxxx, Start: 2014-03-02 14:39:11, End: 2014-03-02 14:39:31''.

\vspace{2pt}\noindent {\bf Audio files}. Almost all the popular instant messaging apps, like WhatsApp, Line, WeChat, Tencent QQ, and KakaoTalk, store the audio files into shared storage directly without any process.

\end{document}